# SPH simulation of fuel drop impact on heated surfaces

Xiufeng Yang [a], Manjil Ray [a], Song-Charng Kong [a,*], Chol-Bum M. Kweon [b]

[a] Department of Mechanical Engineering, Iowa State University, Ames, IA 50011, USA

[b] Propulsion Division, U.S. Army Research Laboratory, APG, MD 21005, USA

* Corresponding author: kong@iastate.edu

**Abstract**

The interaction of liquid drops and heated surfaces is of great importance in many applications. This paper describes a numerical method, based on smoothed particle hydrodynamics (SPH), for simulating n-heptane drop impact on a heated surface. The SPH method uses numerical Lagrangian particles, which obey the laws of fluid dynamics, to describe the fluid flows. By incorporating the Peng-Robinson equation of state, the present SPH method can directly simulate both the liquid and vapor phases and the phase change process between them. The numerical method was validated by two experiments on drop impact on heated surfaces at low impact velocities. The numerical method was then used to predict drop-wall interactions at various temperatures and velocities. The model was able to predict the different outcomes, such as rebound, spread, splash, breakup, and the Leidenfrost phenomenon, consistent with the physical understanding.

**Total length of paper using method 1: 5684**

| | |
|---|---|
| Main text | 2467 |
| Equations | 471 |
| Nomenclature | 0 |
| References | 507 |
| Tables | 0 |
| Figures | 2239 |

# 1. Introduction

The dynamics of drop impact on a solid surface is an important subject to study, as it is commonly encountered in many industrial processes, such as fuel sprays in internal combustion engines, spray cooling or coating of surfaces, ink jet printing, and material processing [1-3]. Numerous studies have dealt with drop impact on heated surfaces in various applications [4-6]. In addition to the impact of a single drop on heated walls, experiments were conducted to characterize the secondary atomization resulting from the simultaneous impact of three parallel liquid drops [7]. The outcome of drop impact depends on both the drop and the wall conditions; it is possible to control the dynamic Leidenfrost temperature by using micro-structured surfaces [8]. An experimental study, using water and fuel droplets, indicates that the wetting behavior is changed by surface topography, wettability, and liquid properties [9]. At high surface temperatures, vaporization becomes dominant at the film boiling regime. The boiling of liquid drops on heated walls near the Leidenfrost point was studied, and three different boiling phenomena were identified, namely, reflective rebound, explosive rebound, and explosive detachment [10]. High-speed color interferometry was used to characterize the air layer thickness and velocity profile between the liquid drop and the wall [11].

Numerous numerical studies on drop-wall interactions have been performed. Most of the numerical studies were based on the mesh-based methods such as the volume of fluid (VOF) method [12-14], level-set (LS) method [15, 16], and coupled VOF-LS method [17]. Challenges arise for these mesh-based methods in tracking the liquid-gas interfaces, especially when the drop splashes and breaks up after impact.

To avoid difficulties in interface tracking, the smoothed particle hydrodynamics (SPH) method was recently formulated to simulate drop-wall interactions at isothermal conditions [18, 19]. The SPH method is a Lagrangian particle method without using a computational mesh. In SPH, a continuous fluid is discretized using SPH particles, which carry physical properties, such as mass, density, pressure, viscosity, and velocity. A smoothing kernel is introduced to connect the neighboring particles. The SPH method can also simulate phase change by incorporating proper equations of state (EOS). Some studies used the non-dimensional van der Waals (vdW) EOS to predict the liquid-vapor equilibrium [20, 21]. Results were qualitatively assessed, since non-dimensional parameters were used in the vdW-EOS. Rapid evaporation and explosive boiling were predicted [22]. However, the existing studies using SPH with vdW-EOS have not dealt with fluids with realistic properties.

This paper describes an SPH method for studying the interaction of n-heptane drops and heated walls. The Peng-Robinson EOS with realistic fuel properties are incorporated into the SPH governing equations. Phase change is predicted, and various outcomes of drop-wall interactions are validated against experimental data.

## 2. Numerical methodology

The governing equations of fluid can be written in the following Lagrangian form.

$$\frac{d\rho}{dt} = -\rho \nabla \cdot \boldsymbol{u} \tag{1}$$

$$\frac{d\boldsymbol{u}}{dt} = \frac{1}{\rho} \nabla \cdot \boldsymbol{S} + \boldsymbol{g} \tag{2}$$

$$\frac{de}{dt} = \frac{1}{\rho} \boldsymbol{S} : \nabla \boldsymbol{u} - \frac{1}{\rho} \nabla \cdot \boldsymbol{q} \tag{3}$$

Here $\rho$ is density, $\boldsymbol{u}$ is velocity, $\boldsymbol{g}$ is the gravitational acceleration, and $e$ is the specific internal energy. The stress tensor $\boldsymbol{S}$ and heat flux vector $\boldsymbol{q}$ are written in the following form

$$\boldsymbol{S} = -p\boldsymbol{I} + \eta(\nabla\boldsymbol{u} + \boldsymbol{u}\nabla) + \left(\zeta - \frac{2}{d}\eta\right)(\nabla \cdot \boldsymbol{u})\boldsymbol{I} \tag{4}$$

$$\boldsymbol{q} = -\kappa \nabla T \tag{5}$$

where $p$ is pressure, $\boldsymbol{I}$ is unit tensor, $\eta$ and $\zeta$ are the coefficients of shear and bulk viscosity, respectively, $\kappa$ is thermal conductivity, and $T$ is temperature.

The governing equations are closed by the following Peng-Robinson equations of state (PR-EOS).

$$p = \frac{RT}{V_m - b} - \frac{a}{V_m^2 + 2bV_m - b^2} \tag{6}$$

$$e = cT - a\rho \tag{7}$$

Here $R$ is the universal gas constant, $a$ and $b$ are parameters, $V_m$ is molar volume. The value of $c$ is chosen such that it could predict the saturated liquid and vapor enthalpies accurately.

In SPH, the value of a function $f$ and its first derivation can be expressed as summations of the neighboring particles.

$$f(\mathbf{r}_a) = \sum_b \frac{m_b}{\rho_b} f(\mathbf{r}_b) W(\mathbf{r}_a - \mathbf{r}_b, h) \tag{8}$$

$$\nabla f(\mathbf{r}_a) = \sum_b \frac{m_b}{\rho_b} f(\mathbf{r}_b) \nabla_a W(\mathbf{r}_a - \mathbf{r}_b, h) \tag{9}$$

The subscripts $a$ and $b$ denote two arbitrary SPH particles, $\mathbf{r}$ is the particle position vector, $m$ is the mass of a particle, $W$ is the kernel function, and $h$ is the smoothing length. $\nabla_a W(\mathbf{r}_a - \mathbf{r}_b, h)$ is the gradient of $W$ at the position of particle $b$ while the coordinate origin is at particle $a$. The SPH form of derivation is not unique, and more details are in literature [23]. The summations are taken over all of the neighboring particles. $W$ determines the domain of neighborhood, while $h$ determines the width of $W$. In this paper, the following kernel function is used [24, 25]

$$W_{ab} = W(\mathbf{r}_a - \mathbf{r}_b, h) = \alpha_d \begin{cases} s^3 - 6s + 6, & 0 \leq s < 1 \\ (2-s)^3, & 1 \leq s < 2 \\ 0, & 2 \leq s \end{cases} \tag{10}$$

where $s = |\mathbf{r}_a - \mathbf{r}_b|/h$, and $\alpha_d$ have the values of $1/(3\pi h^2)$ and $15/(62\pi h^3)$ in two and three dimensions, respectively.

Each particle has its own smoothing length, and the magnitude of $h$ is related to the volume of the particle. In this paper, the following variable smoothing length is used.

$$h = \lambda \left(\frac{m}{\rho}\right)^{1/d} \tag{11}$$

Here $d$ is the spatial dimension. The parameter $\lambda$ is used to control the ratio of smoothing length to particle size. According to Eq. (11), the smoothing length changes with density and may be different for different particles. In order to make sure the interactions between two particles are symmetric, symmetrized kernel function and its gradient are used.

$$\overline{W}_{ab} = \overline{W}_{ba} = \frac{1}{2}[W(\mathbf{r}_a - \mathbf{r}_b, h_a) + W(\mathbf{r}_b - \mathbf{r}_a, h_b)] \tag{12}$$

$$\nabla_a \overline{W}_{ab} = -\nabla_b \overline{W}_{ba} = \frac{1}{2}[\nabla_a W(\mathbf{r}_a - \mathbf{r}_b, h_a) - \nabla_b W(\mathbf{r}_b - \mathbf{r}_a, h_b)] \tag{13}$$

The following can be obtained by applying Eq. (8) to calculate density.

$$\rho_a = \sum_b m_b \overline{W}_{ab} \tag{14}$$

With Eq. (14) for density, the mass equation (1) will be not used in this method. The momentum and energy equations (2) and (3) can be discretized as the following SPH equations. For more details, refer to [20, 21, 24].

$$\frac{d\boldsymbol{u}_a}{dt} = \sum_b m_b \left( \frac{\boldsymbol{S}_a}{\rho_a^2} + \frac{\boldsymbol{S}_b}{\rho_b^2} \right) \cdot \nabla_a \overline{W}_{ab} + \boldsymbol{g}_a \tag{15}$$

$$\frac{de_a}{dt} = \sum_b m_b \left( \frac{\boldsymbol{S}_a}{\rho_a^2} + \frac{\boldsymbol{S}_b}{\rho_b^2} \right) : (\boldsymbol{u}_a - \boldsymbol{u}_b) \nabla_a \overline{W}_{ab} - \sum_b m_b \left( \frac{\boldsymbol{q}_a}{\rho_a^2} + \frac{\boldsymbol{q}_b}{\rho_b^2} \right) \cdot \nabla_a \overline{W}_{ab} \tag{16}$$

## 3. Results and discussions

The experiments of Moita and Moreira [9], using ethanol drops, were first used to test the present model. This validation will provide the evidence that the present SPH method, coupled with PR-EOS, can predict the deformation of a drop when it impinges on a solid surface. Figure 1 shows the comparison of experimental data and SPH simulations using both two- and three-dimensional configurations. Both the measured and predicted spread profiles indicate a decay in the rate of spreading, as expected. A visual inspection of the phenomena (Fig. 2), although qualitative, shows that the simulation can reproduce the experimental observations. After $t = 0.5$ ms, Moita and Moreira [9] reported the gradual formation of a crown from the edge of the liquid film. However, this was not seen in the simulation. Instead, some SPH particles broke away from the edge instead of moving out and up as a thin film. The limited number of SPH particles could be a possible cause of this inconsistency, since there may not be enough SPH particle interactions to represent liquid behaviors after a certain time.

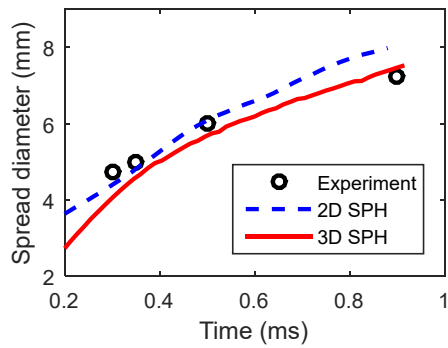

Fig. 1. Spread history of a 2.4-mm ethanol drop impacting on a dry surface at $T_w = 24$ ºC with $U = 3.1$ m/s.

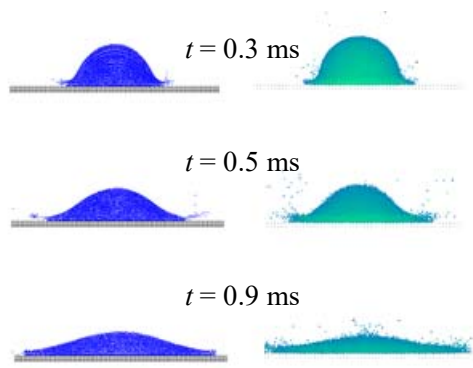

Fig. 2. Side view of 2D (left) and 3D (right) numerical results of a 2.4-mm ethanol drop impact on a dry wall.

After the validation at the room temperature condition, the model was further tested in predicting the impact of n-heptane drops on heated walls. The experiments used n-heptane drops of 1.5 mm diameter, impacting on a stainless-steel surface heated to different temperatures [26]. Figure 3 shows that the numerical results compare well with the experimental data. The spread factor is the ratio of the liquid diameter on the wall to the original drop diameter.

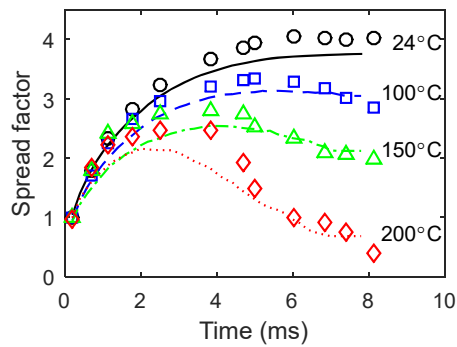

Fig. 3. Spread factor of n-heptane drops ($U$=0.93 m/s) impacting on a stainless-steel surface at different wall temperatures (data points: experimental data [26]; lines: numerical results).

Figure 4 shows the history of drop deformation for different wall temperatures. In the initial 1 ms after impact, the drop spreads rapidly as a thin film. Then the spreading slows down due to the surface tension, viscosity of the liquid, and the friction between the wall and the liquid. Although the outer edge of the drop is retarding, the liquid at the center still has sufficient inertia to keep moving outward, leading to the formation of a cusp. On the outer side of the cusp, the adhesive force between the liquid and the wall will increase the spread, while the surface tension on the inside will pull it back toward the center to restore the drop. The final spread diameter depends on the balance of these two opposing forces. In both the simulation and experimental results (Fig. 3), the spreading rate and the maximum spread factor reduce with increased wall temperature. For wall temperatures above 100°C, the spread diameter reaches its maximum and then reduces because the drop recoils, unlike the 24°C case in which the drop keeps spreading until equilibrium is reached. The following is a possible explanation.

Initially the drop spreads rapidly due to inertia, while the surface tension and viscosity will slow it down. As the inertia is dissipated or converted into surface energy, the cohesive and adhesive forces tend to bring the drop to its final equilibrium position. At lower wall temperatures, the adhesive force between the liquid and surface is higher, and it reduces with increased wall temperature. At higher wall temperatures, the film spreads to a maximum diameter before the inertia is lost, and then the surface tension causes the liquid to recoil, thus reducing the spread diameter. At the Leidenfrost temperature of 200°C and above, the adhesive force between the liquid and the wall disappears because of the formation of a thin layer of vapor. As a result, the liquid completely recoils and rebounds from the surface.

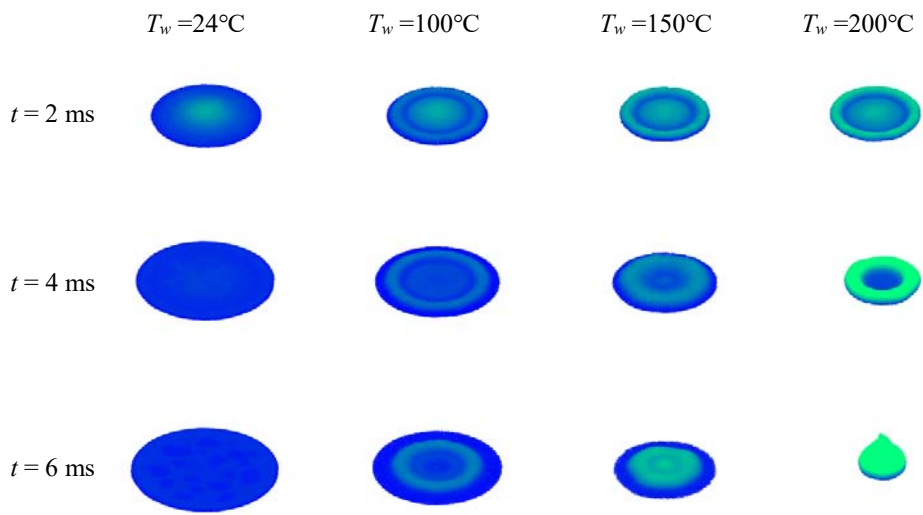

Fig. 4. Impact of a 1.5-mm n-heptane liquid drop on a surface at different temperatures (vapor phase has been removed from the illustration to emphasize the liquid surface structure).

The outcomes of the drop-wall interactions can be divided into two regimes, depending on if the drop disintegrates. At every wall temperature, there is a critical Weber number ($We = \rho U^2 D/\sigma$), beyond which the drop disintegrates, producing secondary droplets. The Weber number is evaluated using the density and surface tension at room temperature, which is the initial liquid drop temperature. To determine the critical $We$, simulations were conducted using different drop impact velocities for a given wall temperature. The smallest $We$ that leads to the ejection of secondary droplets is noted as the critical $We$. Results show that, for wall temperatures lower than the Leidenfrost point, the critical $We$ decreases with increased wall temperature, as shown in Fig. 5. Above the Leidenfrost point, the critical $We$ remains the same.

Below the Leidenfrost temperature, at high $We$, a liquid film is formed with numerous tiny droplets ejected upon impact. At the same impact velocity, more secondary droplets are generated at higher wall temperatures. Consequently, the critical $We$ is lower at elevated temperatures (Fig. 5). Beyond the Leidenfrost point, the liquid disintegrates into smaller drops at high $We$ (Fig. 6). At the critical $We$ of 100, the drop breaks up during the recoil period, a phenomenon known as receding breakup. At a slightly higher $We$ of 115, the drop disintegrates before the start of the recoil period and most of the resulting droplets reside on the surface. This outcome is called drop disintegration. At an even higher $We$ of 204, the drop breaks up into even smaller droplets, which are ejected upwards from the surface. This signifies the so-called splashing regime. Above the Leidenfrost temperature, the critical $We$ does not depend on an increase in wall temperature (Fig. 5). This agrees with the understanding that in the film boiling regime the impact outcome depends only on the impact velocity, since a vapor film is formed to prevent the direct contact of liquid and wall.

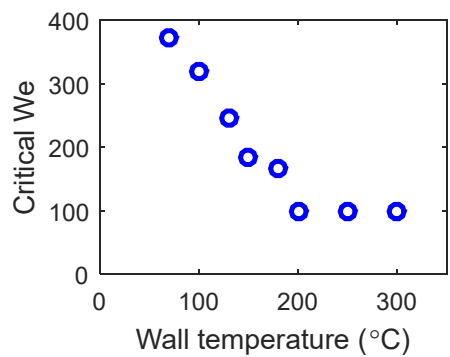

Fig. 5. Predicted critical Weber number as a function of temperature (1.5 mm n-heptane drop).

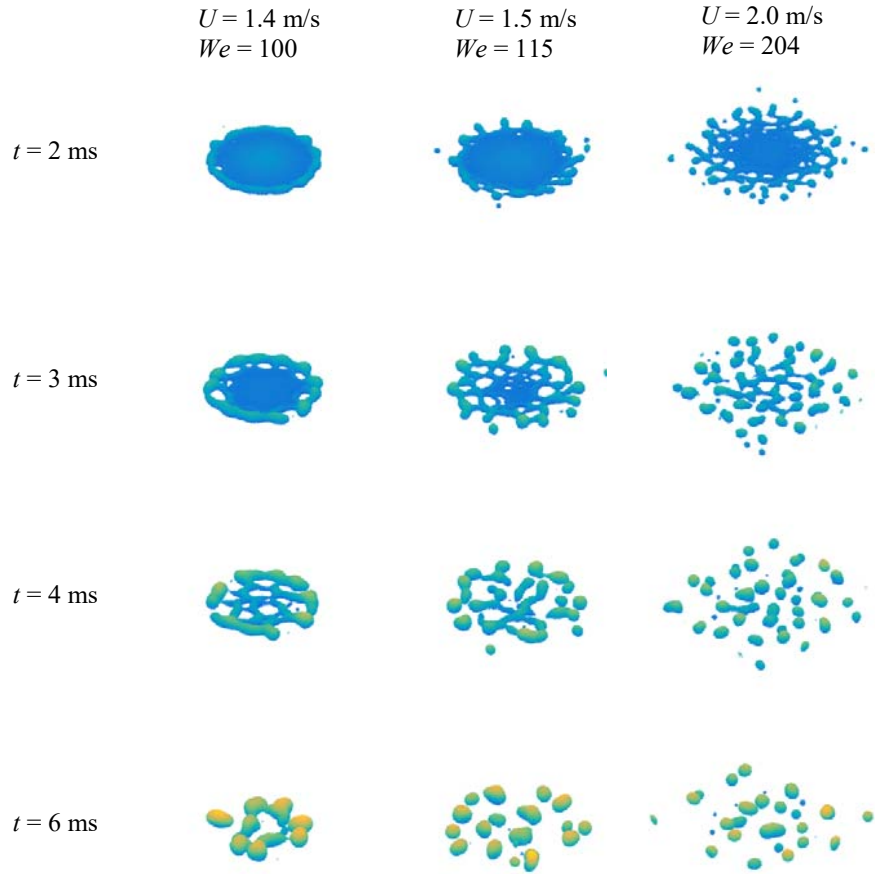

Fig. 6. Three regimes of drop breakup for wall temperature ($T_w$ = 350°C) above the Leidenfrost point (from left to right: receding breakup, disintegration, splashing).

Earlier experiments using water drops have shown that at wall temperatures above the Leidenfrost point, the drop rebound velocity did not increase monotonically with the impact velocity [27]. Instead, at low $We$, the rebound velocity increases with the impact velocity and reaches a maximum, beyond which the drop flattens and the rebound velocity decreases. Under the conditions studied [27], the drop disintegrates upon impact for $We > 80$, resulting in a wall jet, if the incident angle is not normal to the wall. To further test the present model, simulations were performed using n-heptane drops of 1.5-mm diameter impacting vertically on a solid surface of 350°C. Figure 7 shows the history of drop movement at various impact conditions, and Fig. 8 shows the relationship between the impact *We* and rebound *We*. For the 1.5-mm diameter drop, the maximum rebound *We* of 5.4 corresponds to an impact velocity of 0.88 m/s. The trend observed in the experiments [27] was captured by the present model.

As the drop spreads, the liquid moves away from the center, converting the kinetic energy into viscous and surface energies. When the velocity is increased, the drop deforms to a greater extent and loses more energy due to the viscous dissipation. As a result, the hysteresis (i.e. the inability to revert back to its original shape) increases, which is manifested as an increased amplitude of oscillation. Later, as the drop rebounds from the wall, it eventually regains the spherical shape because of the surface tension. When the impact velocity reaches 1.4 m/s (*We* = 100 for the 1.5-mm diameter drop), the drop no longer rebounds cleanly. This impact condition represents a limit and signifies the critical level of hysteresis required to prevent rebound. The rebound velocity at this condition is approximately zero for the 1.5-mm diameter drop. Beyond this point, drop disintegration intensifies, and the impact is considered to be outside the rebound regime. Overall, as indicated in Fig. 8, between very low incident velocities and the critical *We* lies the inflection point (*We* ~ 40), corresponding to an impact velocity of 0.88 m/s at which the rebound *We* is the highest. Prior to reaching the reflection point, the impact will result in an elastic rebound [27]. Between the inflection point and the critical *We*, the drop flattens significantly and the rebound velocity keeps decreasing. Beyond the critical *We*, the drop disintegrates upon impact, which will result in a wall jet from the point of impact.

To study the variability of this trend, simulations were also performed using a 0.5-mm diameter n-heptane drop. The rebound characteristics were nearly the same, as shown in Fig. 8, except that the critical *We* is higher (*We* = 154) for the 0.5-mm diameter drop, likely because the drop is already small and is less prone to breakup.

In perspective, this paper has presented a computational tool that is able to predict the details of drop-wall interactions. The numerical model considers the fundamental fluid dynamics equations, coupled with the PR-EOS with realistic thermodynamic properties. This computational tool can be extended to simulate other fuels and predict the outcomes of drop-wall interactions at practical operating conditions. The numerical results can be further analyzed for developing engineering models for spray combustion application.

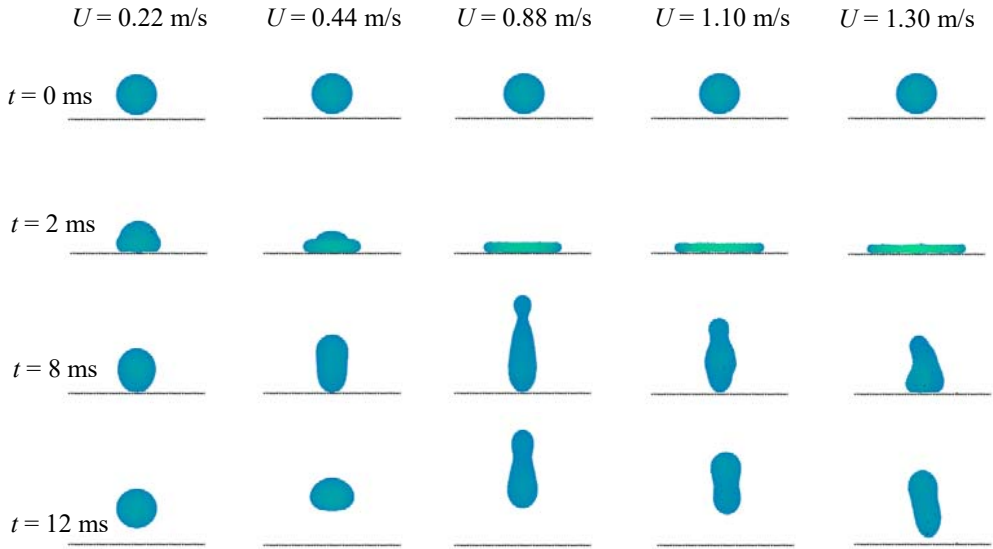

Fig. 7. History of liquid drop deformation and rebound from a hot wall ($T_w = 350°C$) at different impact conditions (1.5 mm n-heptane drop).

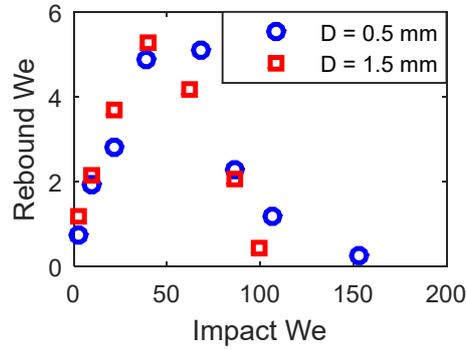

Fig. 8. Rebound $We$ vs impact $We$ for n-heptane drops ($D = 0.5$ mm and 1.5 mm; $T_w = 350°C$).

## 4. Conclusions

An SPH method for simulating the details of drop impact on heated surfaces is presented in this paper. The PR-EOS is used for both the liquid and vapor phases. Phase change between the liquid and vapor phases is simulated directly without the need to use a vaporization model. The present SPH method is validated and applied to predict the impact of both ethanol and n-heptane drops on heated surfaces. The present model is able to predict the different impact regimes and the relationship between impact velocity and rebound velocity. Numerical results show that the critical Weber number, at which secondary droplets are generated upon impact, decreases with the increase in the wall temperature. The present model can be further used to predict the detailed outcomes of drop-wall interactions at combustion-relevant conditions for developing engineering models for use in spray combustion simulations.

## Acknowledgement

Support from the U.S. Army Research Laboratory is acknowledged.